\documentstyle{l-aa}
\def\be{\begin{equation}}
\def\ee{\end{equation}}
\def\bea{\begin{eqnarray}}
\def\eea{\end{eqnarray}}
\clearpage
\newpage
\thesaurus{}

\title{On the deprojection of triaxial galaxies with
St\"ackel potentials}

\author{A.~Mathieu and H.~Dejonghe}

\institute{Universiteit Gent, Sterrenkundig Observatorium,
	Krijgslaan 281, B--9000 Gent, Belgium}

\date{Received date; accepted date}

\offprints{A. Mathieu }

\begin{document}
\maketitle

% -----------------------------------------------------------------------
\begin{abstract}
% -----------------------------------------------------------------------

A family of triaxial St\"ackel potential-density pairs is introduced.
With the help of a Quadratic Programming method, a linear combination
of potential-density pairs of this family which fits a given projected
density distribution can be built.  This deprojection strategy can be
used to model the potentials of triaxial elliptical galaxies with or
without dark halos. Besides, we show that the expressions for the
St\"ackel triaxial density and potential are considerably simplified
when expressed in terms of divided differences, which are convenient
numerically. We present an example of triaxial deprojection for the
galaxy NGC~5128 whose photometry follows the de Vaucouleurs law.
 
% ----------------------------------------------------------------------- 
\end{abstract}
% -----------------------------------------------------------------------


% -----------------------------------------------------------------------
\section{Introduction}
% -----------------------------------------------------------------------

As soon as one realized that elliptical galaxies were not spheroids
flattened by rotation, a growing list of unexpected photometric and
kinematical observations has led astronomers to develop a more
elaborate picture of these objects.  While triaxiality provides a
plausible explanation of these new findings, it also raises a number
of questions such as the determination of the intrinsic shapes of
these stellar systems.

The deprojection problem consists in finding the intrinsic mass (or
light) density from its projection onto the plane of the sky.  It is
well known that the knowledge of the surface brightness distribution
does not define uniquely the intrinsic three-dimensional light
distribution and the orientation of the galaxy. Even in the
axisymmetric case the deprojection is degenerate: one can construct
many density distributions that project to the same photometric
distribution (see e.g. Gerhard \& Binney 1995). Features such as the
presence of lanes or disks in some galaxies have been exploited to
further constrain the problem.  Recently Statler (1994) proposed a new
method for constraining the intrinsic shapes of elliptical galaxies
using information on their apparent shapes and velocity fields.

In this paper, we shall describe a technique that determines an
intrinsic mass density distribution of a galaxy consistent with a
given projected mass density. Similarly, the method allows to
determine the intrinsic light distribution of a galaxy given its
surface brightness distribution.  This is the first step towards a
triaxial dynamical model.  St\"ackel potentials are particularly
suited to this kind of modeling since the investigation of the
dynamics is essentially analytical.  A St\"ackel triaxial mass model
can be constructed by specifying a density profile along the short
axis and an ellipsoidal coordinate system. Simple examples of such
models are given in de Zeeuw, Peletier \& Franx (1986). However, the
determination of the potential generally requires the evaluation of
one-dimensional quadratures.  Another approach consists in specifying
ellipsoidal coordinates and a simple form for the St\"ackel potential.
The mass density can be calculated by means of the generalized
Kuzmin's formula (de Zeeuw 1985b) which involves straightforward
derivatives of a one-dimensional function.  We shall use this method
in order to have a simple analytical expression for the potential.

The basic properties of St\"ackel models are studied in Sect.~2.
Section~3 presents a family of St\"ackel potential-density pairs that
can be used as building blocks for the construction of triaxial mass
models. Examples of projected and spatial density distributions are
shown and then we describe the deprojection method. In Sect.~4, we
present a fit to a triaxial modified Hubble model using these
potential-density pairs. Section~5 deals with the application of the
deprojection method to the elliptical galaxy NGC~5128 (Centaurus~A)
whose photometry follows the de Vaucouleurs law. Our conclusions are
given in Sect.~6.

% -----------------------------------------------------------------------
\section{Triaxial St\"ackel models}
% -----------------------------------------------------------------------

A sequence of potential-density pairs can be generated by repeated
application of an operator upon a given potential-density pair. Such a
method has been used by de Zeeuw and Pfenniger (1988, hereafter
referred to as ZP88) to build a class of triaxial potential-density
pairs ($V_n$,\,$\rho_n$), amongst which the family ($V_2$,\,$\rho_2$)
is of St\"ackel form.  In the next subsection, we present some
properties of the St\"ackel models (for a detailed discussion see de
Zeeuw 1985a, ZP88).

\subsection{General properties}

Let ($x$,\,$y$,\,$z$) be cartesian coordinates and
($\lambda$,\,$\mu$,\,$\nu$) be ellipsoidal coordinates defined
according to de Zeeuw (1985a).  In these coordinates, the St\"ackel
potential, for which the equations of motion separate, has the general
form
\be
V(\lambda,\mu,\nu) = g_\lambda F(\lambda) + g_\mu F(\mu) + g_\nu
F(\nu),
\label{potstackdef}
\ee
where $F$ is an arbitrary function and
\be
g_\lambda  =
\frac{(\lambda+\alpha)(\lambda+\beta)}{(\lambda-\mu)(\lambda-\nu)}
\quad (cyc).
\label{gtaudef}
\ee
$\alpha$, $\beta$ and $\gamma$ are negative constants and
\be
-\gamma \leq \nu \leq -\beta \leq \mu \leq -\alpha \leq \lambda.
\ee
Similar expressions for $g_\mu$ and $g_\nu$ are found by a cyclic
permutation $\lambda \! \rightarrow \! \mu \!
\rightarrow \! \nu \! \rightarrow \! \lambda$.  Furthermore, we have
$0\leq~g_\lambda\,,\,g_\mu\,,\,g_\nu\leq~1$ and 
\be
g_\lambda+g_\mu+g_\nu=1.
\label{sumgtau}
\ee
The triaxial density follows from Poisson's equation and can be
written (see ZP88) as
\be
\rho(\lambda,\mu,\nu) = g_\lambda^2 \Psi'(\lambda) + g_\mu^2
\Psi'(\mu) + g_\nu^ 2\Psi'(\nu) \,+ 
\label{densstackdef}
\ee
\begin{displaymath}
\quad \quad \quad 2 g_\lambda g_\mu \Psi[\lambda,\mu]
+ 2 g_\mu g_\nu \Psi[\mu,\nu]
+ 2 g_\nu g_\lambda \Psi[\nu,\lambda],
\end{displaymath}
where $\Psi[\tau_1,\tau_2]$ denotes the first order divided
difference of the function $\Psi(\tau)$ i.e.
\be
\Psi[\tau_1,\tau_2] = \frac{\Psi(\tau_1) -
\Psi(\tau_2)}{\tau_1-\tau_2}
\ee
and $\Psi'(\tau)$ is the derivative $\Psi[\tau,\tau]$. The relation
that connects $\Psi(\tau)$ and $F(\tau)$ is (ZP88):
\be
2\pi G \Psi(\tau) = 2(\tau+\gamma)F'(\tau) - F(\tau) \,+ 
\ee
\begin{displaymath}
\quad (\tau+\gamma) 
\left(
\frac{ F(\tau)-F(-\alpha) }{\tau+\alpha} + 
\frac{ F(\tau)-F(-\beta) }{\tau+\beta} 
\right).
\end{displaymath}
Given a function $F(\tau)$, one can calculate the St\"ackel potential
with Eq.~(\ref{potstackdef}), and the density using
Eq.~(\ref{densstackdef}).  The density profile along the $z$-axis is
the function $\Psi'(\tau)$ with $\tau = z^2-\gamma$. The triaxial
St\"ackel density is fully determined by the specification of the
density on the short $z$-axis; it can be written as a weighted sum of
the density at six particular points of the $z$-axis (de Zeeuw 1985b).

\subsection{Expressions for the St\"ackel potential and density in
terms of divided differences}

The use of divided differences considerably simplifies the formulas of
the St\"ackel triaxial density and potential. The divided difference
of order $(n-1)$ of the function $U(\tau)$ is a function of divided
differences of order $(n-2)$ and is given by
\be
U[\tau_1,\tau_2,\cdots,\tau_n] = \frac{
U[\tau_1,\tau_3,\cdots,\tau_n] -
U[\tau_2,\tau_3,\cdots,\tau_n]}{\tau_1 - \tau_2}.
\ee
The second order divided difference of a general function $U(\tau)$
can be written explicitly as
\be
U[\lambda,\mu,\nu] = \frac{U(\lambda)}{(\lambda-\mu)(\lambda-\nu)} +
(cyc).
\label{defdivdif3}
\ee
With Eqs.~(\ref{gtaudef}) and (\ref{potstackdef}), we can express the
potential $V$ as the second order divided difference
\be
V(\lambda,\mu,\nu) = U[\lambda,\mu,\nu]
\label{potdivdif}
\ee
with
\be
U(\tau) = (\tau+\alpha)(\tau+\beta)F(\tau).
\ee

The density $\rho(\lambda,\mu,\nu)$ can be written in terms of divided
differences using either the function $\Psi(\tau)$ whose derivative is
the density profile along the $z$-axis, or using the basis function
$F(\tau)$ that defines the potential $V$. The detailed proofs are
given in Appendix B and Appendix C respectively.  To establish these
results, we shall use the potential-density pair $(\rho_1,V_1)$
introduced in ZP88. Its main properties are given in Appendix A. In
this subsection, we only present the relevant results.

The triaxial St\"ackel density $\rho(\lambda,\mu,\nu)$ can be
expressed as the $5^{\rm th}$-order divided difference of the function
$H(\tau)$
\be
\rho(\lambda,\mu,\nu) = H[\lambda,\mu,\nu,\lambda,\mu,\nu]
\label{densdivdifbis}
\ee
with 
\be
H(\tau) = (\tau+\alpha)^2(\tau+\beta)^2\Psi(\tau).
\label{deffuncgtaubis}
\ee

It can also be written as the $5^{\rm th}$-order divided difference of
the function $R(\tau)$
\be
\rho(\lambda,\mu,\nu) = R[\lambda,\mu,\nu,\lambda,\mu,\nu]
\label{densdivdifr2bis}
\ee
with
\be
\pi G R(\tau) = |a(\tau)|^{\frac{3}{2}} \frac{d}{d
\tau} \left[ \frac{\sqrt{|a(\tau)|}}{\tau+\gamma} F(\tau) \right],
\label{funcrtaubis}
\ee
where
\be
a(\tau) = (\tau+\alpha)(\tau+\beta)(\tau+\gamma).
% \label{defatau}
\ee

% -----------------------------------------------------------------------
\section{The deprojection strategy}
%------------------------------------------------------------------------

\subsection{Basis component}

A St\"ackel triaxial mass model is completely determined by the choice
of ellipsoidal coordinates and a function $F(\tau)$.  So as to have
simple analytical expressions for the potential and the density, we
choose $F(\tau)$ to be an elementary function of $\tau$.

The spherical H\'enon's isochrone model (1959) is often used in the
construction of models of stellar systems. More realistic models that
account for the flattening of the potential can be produced with
axisymmetric generalizations (see Dejonghe \& de Zeeuw 1988; Evans, de
Zeeuw \& Lynden-Bell 1990). Axisymmetric St\"ackel models that reduce
to the isochrone in the spherical limit can be constructed using
\be
F(\tau) = -\frac{GM}{\sqrt{-\alpha} + \sqrt{\tau}}\ .
\ee
The associated potential is the thoroughly studied Kuzmin-Kutuzov
potential (see e.g. Kuzmin 1956, Kuzmin \& Kutuzov 1962, Dejonghe \&
de Zeeuw 1988, Batsleer \& Dejonghe 1993). Using the above function
$F(\tau)$, a triaxial generalization can also be calculated with
Eqs.~(\ref{potstackdef}) and (\ref{densstackdef}) or, alternatively,
Eqs.~(\ref{densdivdifr2bis}) and (\ref{funcrtaubis}). Expressions for
the density and the potential are given in ZP88.

We choose a three-parameter basis function of the form
\be
F(\tau) = -\frac{GM}{ {(d + \tau^p)}^s}
\label{basiscomp}
\ee
where $d$, $p$ and $s$ are real parameters. The triaxial isochrone has
$p=0.5$, $s=1$ and $d=\sqrt{-\alpha}$.

\subsection{Some examples}

Kinematic studies of elliptical galaxies have revealed a very diverse
and complicated nature of these objects. There has been marked
improvement in the photometric observations with the advent of CCDs
and it appears that ellipticals also span a wide range of photometric
properties. The isophotal shape is mainly elliptical, but departures
from perfect ellipses are often detected, as well as isophote twists
(see e.g. Bender, D\"{o}bereiner \& M\"{o}llenhoff 1988).

The projected mass densities of the basis components of the form
(\ref{basiscomp}) have elliptical surface isodensities that can
exhibit deviations from pure ellipses such as boxiness and diskiness.
Thus these components may be relevant to approximate the photometry of
elliptical galaxies.  It is well known that St\"ackel potentials
cannot produce projected densities with isophote twists (Franx 1988),
but this is often a second order effect compared to the ellipticity
variation.

In this section, we present three examples of spatial and projected
density distributions for a normal elliptical
(Fig.~\ref{contourelliptic}), a discy (Fig.~\ref{contourdiscy}) and a
boxy (Fig.~\ref{contourboxy}) models. The units are arbitrary.  The
intrinsic long, intermediate and short axes are denoted $x$, $y$ and
$z$ respectively. The models have
($-\alpha$\,,\,$-\beta$\,,\,$-\gamma$) = (7.3\,,\,3.6\,,\,0.8).  The
parameters ($d$\,,~$p$\,,~$s$) of the components are
(5\,,\,0.7\,,\,0.7) for the elliptical model, (4\,,\,0.5\,,\,1) for
the discy model and (5\,,\,0.93\,,\,0.44) for the boxy model.

\begin{figure}
\vspace*{9cm}
\special{hscale=22 vscale=22 hoffset=-5 voffset=80 
 hsize=200 vsize=260 angle=0 psfile="fig1a.ps"}
\special{hscale=22 vscale=22 hoffset=120 voffset=80 
 hsize=300 vsize=260 angle=0 psfile="fig1b.ps"}
\special{hscale=22 vscale=22 hoffset=-5 voffset=-50 
 hsize=200 vsize=250 angle=0 psfile="fig1c.ps"}
\special{hscale=22 vscale=22 hoffset=120 voffset=-50 
 hsize=300 vsize=250 angle=0 psfile="fig1d.ps"}
\caption{Contour map of the spatial mass density in the planes (top
left) $x=0$, (top right) $y=0$ and (bottom left) $z=0$ for an
elliptical model. Bottom right: contour map of the projected mass
density. The contour step is 0.65 magnitude}
\label{contourelliptic}
\end{figure}

\begin{figure}
\vspace*{9cm}
\special{hscale=22 vscale=22 hoffset=-5 voffset=80 
 hsize=200 vsize=260 angle=0 psfile="fig2a.ps"}
\special{hscale=22 vscale=22 hoffset=120 voffset=80 
 hsize=300 vsize=260 angle=0 psfile="fig2b.ps"}
\special{hscale=22 vscale=22 hoffset=-5 voffset=-50 
 hsize=200 vsize=250 angle=0 psfile="fig2c.ps"}
\special{hscale=22 vscale=22 hoffset=120 voffset=-50 
 hsize=300 vsize=250 angle=0 psfile="fig2d.ps"} 
\caption{Contour map of the spatial mass density in the planes (top
left) $x=0$, (top right) $y=0$ and (bottom left) $z=0$ for a discy
model.  Bottom right: contour map of the projected mass density. The
contour step is 0.6 magnitude}
\label{contourdiscy}
\end{figure}

\begin{figure}
\vspace*{9cm}
\special{hscale=22 vscale=22 hoffset=-5 voffset=80 
 hsize=200 vsize=260 angle=0 psfile="fig3a.ps"}
\special{hscale=22 vscale=22 hoffset=120 voffset=80 
 hsize=300 vsize=260 angle=0 psfile="fig3b.ps"}
\special{hscale=22 vscale=22 hoffset=-5 voffset=-50 
 hsize=200 vsize=250 angle=0 psfile="fig3c.ps"}
\special{hscale=22 vscale=22 hoffset=120 voffset=-50 
 hsize=300 vsize=250 angle=0 psfile="fig3d.ps"}
\caption{Contour map of the spatial mass density in the planes (top
left) $x=0$, (top right) $y=0$ and (bottom left) $z=0$ for a boxy
model. Bottom right: contour map of the projected mass density. The
contour step is 0.6 magnitude}
\label{contourboxy}
\end{figure}

\subsection{The method}

We use a Quadratic Programming (QP) method (Dejonghe 1989) to fit to a
given photometric data set a linear combination of basis functions
with the constraint that the density must be positive.  Given a
function $F(\tau)$ of the form (\ref{basiscomp}), the calculation of
the St\"ackel potential and density is straightforward as it involves
only the evaluation of elementary functions. These basis components
can produce physical density distributions (i.e.  positive everywhere
in real space) as well as unphysical ones, depending on the choice of
the ellipsoidal coordinates and the parameters $d$, $p$ and $s$.
Generally, the projected density distribution has to be calculated
numerically.

As showed by Kuzmin (1956) for axisymmetric mass models and
subsequently generalized for triaxial models by de Zeeuw (1985b), the
so-called Kuzmin's theorem states that a triaxial mass density built
with a St\"ackel gravitational potential is everywhere positive if the
density on the $z$-axis is positive. Therefore, the positivity check
of the triaxial density $\rho(\lambda,\mu,\nu)$ reduces to a
one-dimensional check on the density along the $z$-axis.

% -----------------------------------------------------------------------
\section{Triaxial modified Hubble model}
% -----------------------------------------------------------------------

The de Vaucouleurs and the Hubble laws are among the most commonly
used fitting profiles for the photometry of ellipticals.  The density
radial profile of a modified Hubble model $\rho\sim (1+r^2)^{-3/2}$
decreases proportional to $r^{-3}$ at large radii.  A separable
triaxial generalization of the modified Hubble model has been studied
by de Zeeuw, Peletier \& Franx (1986).  The density along the $z$-axis
for a modified Hubble model with core radius $c$ can be written
\be
\Psi'(z) = \frac{\rho_0 c^3}{(z^2+c^2)^{3/2}}.
\ee
By choosing $\gamma=-c^2$ and using $z^2 = \tau+\gamma$ on the
$z$-axis, it becomes
\be
\Psi'(\tau) = \rho_0 {\left(\frac{-\gamma}{\tau}\right)}^{3/2}.
\ee
The function $\Psi(\tau)$ is the primitive 
\be
\Psi(\tau) = \int_{-\gamma}^{\tau} \Psi'(\sigma) d\sigma = 2 \rho_0
\gamma \left( \sqrt{\frac{-\gamma}{\tau}}-1 \right).
\ee
The triaxial density can be calculated with Eqs.~(\ref{densdivdifbis})
and (\ref{deffuncgtaubis}).

Using basis components of the form (\ref{basiscomp}), we can fit the
density on the $z$-axis $\Psi'(\tau)$ of a modified Hubble model. As
the triaxial density is completely determined by the ellipsoidal
coordinate system and the function $\Psi'(\tau)$, we can therefore
produce a fit to a triaxial Hubble model. A wide range of triaxiality
is allowed through the choice of the ellipsoidal coordinates.  In
Fig.~\ref{conthubblephot}, we present the spatial and projected mass
densities of our best fitting model with
($-\alpha$\,,\,$-\beta$\,,\,$-\gamma$) = (4\,,\,2.6\,,\,1). The model
is fully triaxial, with a triaxiality parameter $T$ defined as $T =
(A^2-B^2)/(A^2-C^2)$ (with $A$, $B$ and $C$ the long, intermediate and
short axis lengths of the density) of $\sim0.5$.  The relative
difference between the function $\Psi'(\tau)$ and the fit is smaller
than 1\,\% out to $z\sim40\,c $.

\begin{figure}
\vspace*{9cm}
\special{hscale=22 vscale=22 hoffset=-5 voffset=80 
hsize=200 vsize=260 angle=0 psfile="fig4a.ps"}
\special{hscale=22 vscale=22 hoffset=120 voffset=80 
hsize=300 vsize=260 angle=0 psfile="fig4b.ps"}
\special{hscale=22 vscale=22 hoffset=-5 voffset=-50 
hsize=200 vsize=250 angle=0 psfile="fig4c.ps"}
\special{hscale=22 vscale=22 hoffset=120 voffset=-50 
hsize=300 vsize=250 angle=0 psfile="fig4d.ps"}
\caption{ 
Contour map of the spatial mass density in the planes (top left)
$x=0$, (top right) $y=0$ and (bottom left) $z=0$ for the best fitting
model to a triaxial modified Hubble model. Bottom right: contour map
of the projected mass density (solid line) of the fit. The contour
step is 0.5 magnitude. The dashed line is the projected mass density
of the modified Hubble model. The direction of projection is the
$y$-axis}
\label{conthubblephot}
\end{figure}

% -----------------------------------------------------------------------
\section{Deprojection of a de Vaucouleurs photometry}
% -----------------------------------------------------------------------

\begin{figure}
\vspace*{9cm}
\special{hscale=22 vscale=22 hoffset=-5 voffset=80 
hsize=200 vsize=260 angle=0 psfile="fig5a.ps"}
\special{hscale=22 vscale=22 hoffset=120 voffset=80 
hsize=300 vsize=260 angle=0 psfile="fig5b.ps"}
\special{hscale=22 vscale=22 hoffset=-5 voffset=-50 
hsize=200 vsize=250 angle=0 psfile="fig5c.ps"}
\special{hscale=22 vscale=22 hoffset=120 voffset=-50 
hsize=300 vsize=250 angle=0 psfile="fig5d.ps"}
\caption{ 
Contour map of the spatial light density in the planes (top left)
$x=0$, (top right) $y=0$ and (bottom left) $z=0$ for the best fit
model to Cen~A's photometry. Bottom right: contour map of the
projected light density (solid line). The dashed line is the de
Vaucouleurs photometry model. The contour step is 0.5 magnitude}
\label{contcenaphot}
\end{figure}

Centaurus~A (NGC~5128) is a giant elliptical galaxy with a conspicuous
dust lane lying along its photometric minor axis.  A photometric study
of Cen~A by Dufour et~al. (1979) from photographic plates showed that
the light distribution of Cen~A follows the de Vaucouleurs law at
radial distances from 2 arcmin to 8 arcmin.  Furthermore, they found
that the $V$ surface brightness distribution at radii from 4 arcmin up
to 8 arcmin is consistent with Cen~A being an E2 galaxy. The isophotes
are elliptical in shape and are quite round at the center but become
more flattened with increasing radius. The ellipticity defined as
$\epsilon=1-b/a$ (with $a$ and $b$ the apparent major and minor axis
lengths) increases from $\epsilon=0.07$ at $r=2.6$ arcmin to
$\epsilon=0.26$ at $r=9$ arcmin.  Following Hui et~al. (1993), we
adopt a distance of 3.5~Mpc to Cen~A, so that 1 arcmin corresponds to
1.02~kpc. According to Dufour et~al.  (1979), the effective radius of
the de Vaucouleurs law is $r_e=5.18$~kpc.

We build a model for the photometry of Cen~A that follows the de
Vaucouleurs law and reproduces the observed flattening of the light
distribution out to 8 arcmin.  Beyond this radius, we choose a
constant value of the ellipticity $\epsilon=0.26$ and a wide range of
mass density profiles is allowed that can account for a possible increase
of the mass-to-light ratio with radius. We adopt the observer's
viewing direction determined by Hui et~al. (1995).

We use the QP method described in Sect.~3 to fit a St\"ackel model to
the de Vaucouleurs photometry of Cen~A.  Given ellipsoidal
coordinates, we find that models ranging from quite oblate to fully
triaxial can fit the data.  In general, our QP models consist of less
than $\sim 10$ components. Adding new components does not improve the
fit significantly.  We present a model with a triaxiality parameter
$T$ of $\sim0.5$.  Contours of the spatial density in the principal
planes are shown in Fig.~\ref{contcenaphot}. The contours of the
projected density (solid line) of the QP best fit are compared to the
contours of the de Vaucouleurs photometry model (dashed line).  The
residual differences of 92\,\% of the pixels inside the region limited
by the isophote of major axis length of 35 arcmin are smaller than
0.1~magnitude.  Figure~\ref{profileDPMcena} shows the profiles of the
QP model and the de Vaucouleurs photometry model along the photometric
major and minor axes and the apparent axis ratio as a function of
radius for the two models.  The components allow to reproduce the de
Vaucouleurs law in the range $\sim0.5\, r_e$ to $\sim6
\,r_e$. Taking into account the fact that the mass-to-light ratio is
likely to vary a bit with radius in this galaxy, there is no
particular need to attempt a better fit if a deprojected mass model
and hence potential are desired. The fit yields a sufficient
approximation of the galaxy's potential. In the inner region ($r<0.5\,
r_e$), the fit is less steep than the de Vaucouleurs law (see
Fig.~\ref{profileDPMcena}). Dynamical models produced with this
potential may depend on three integrals, which will be only
approximative in the very center.

\begin{figure}
\vspace*{13cm}
\special{hscale=38 vscale=38 hoffset=0 voffset=150
hsize=300 vsize=850 angle=0 psfile="fig6a.ps"}
\special{hscale=38 vscale=38 hoffset=0 voffset=-30 
hsize=300 vsize=850 angle=0 psfile="fig6b.ps"}
\caption{ 
Top: profiles of the projected density of our QP best fit model to
Cen~A's photometry along the photometric major (bottom solid line) and
minor (top solid line) axes. The dashed lines are the corresponding
profiles of the de Vaucouleurs photometry model. Bottom: apparent axis
ratio as a function of radius for the QP model (solid line) and for
the de Vaucouleurs photometry model (dashed line)}
\label{profileDPMcena}
\end{figure}

% -----------------------------------------------------------------------
\section{Conclusions}
% -----------------------------------------------------------------------

A family of triaxial St\"ackel potential-density pairs is presented.
It includes as a special case the triaxial generalization of H\'enon's
isochrone in ellipsoidal coordinates. This family allows the
construction of triaxial mass (or light) models of galaxies with or
without dark halos.  A large variety of intrinsic shapes is provided
by the choice of ellipsoidal coordinates and by the components. This
diversity is also reflected in the projected densities which show
elliptical, box-like or disc-like isophotes with ellipticities
changing as a function of radius.

These potential-density pairs can be used as building blocks for
realistic St\"ackel models of triaxial potentials in elliptical
galaxies.  This is first tested with the elliptical galaxy Centaurus~A
(NGC~5128) whose kinematics exhibits unambiguous signatures of
triaxiality.  Using a Quadratic Programming method, we find a linear
combination of density distributions that fits a model of the surface
brightness of this E2 galaxy with a total density which is positive
everywhere.  The de Vaucouleurs photometry model of Cen~A is well
reproduced with these components. Mass-to-light ratio variations could
be included in the projected distribution and the same method would
then produce spatial mass models with dark matter.  It suggests that
St\"ackel mass models may be relevant to the description of galactic
potentials which represents a first step towards St\"ackel triaxial
dynamical models of stellar systems.  Abel components propounded by
Dejonghe \& Laurent (1991) may prove powerful in this context; work
along this line was first carried out by Dejonghe (1992), subsequently
by Zeilinger et~al.  (1993) and further investigation is in progress.

\begin{acknowledgements}
We thank the referee E.~Emsellem for his critical comments which
helped us improve the presentation.
\end{acknowledgements}

% -----------------------------------------------------------------------
\section*{\bf Appendices}
%-----------------------------------------------------------------------

\appendix

\section{The potential-density pair $(\rho_1,V_1)$}

Let the potential $V_1$ be of the general form
\be
V_1(\lambda,\mu,\nu) = F_1(\lambda) + F_1(\mu) + F_1(\nu).
\label{defv1}
\ee
The Laplace operator $\nabla^2$ in ellipsoidal coordinates is (see
e.g. de Zeeuw 1985b, Eq.~(20))
\be
\nabla^2 = \nabla^2_\lambda + \nabla^2_\mu + \nabla^2_\nu
\ee
with
\be
\nabla^2_\lambda = \frac{2}{(\lambda-\mu)(\lambda-\nu)}\left(
2a(\lambda)\frac{\partial^2}{\partial\lambda^2} +
a'(\lambda)\frac{\partial}{\partial\lambda}\right) \ (cyc),
\label{defnabla}
\ee
where
\be
a(\tau) = (\tau+\alpha)(\tau+\beta)(\tau+\gamma).
\label{defatau}
\ee
The density $\rho_1$ that follows from Poisson's equation can be
written as (Eq.~(3.9) in ZP88)
\be
\rho_1(\lambda,\mu,\nu) = g_\lambda \Psi_1(\lambda) + (cyc).
\ee
The relation between $\Psi_1(\tau)$ and $F_1(\tau)$ is written
explicitly in ZP88 (Eq.~(3.7)), as well as the relations that connect
the St\"ackel pair $(\rho,V)$ and the pair $(\rho_1,V_1)$. In
particular, we have
\be
\rho = ({\cal A}+{\cal I}) \rho_1,
\label{relrhorho1}
\ee
\be
F(\tau) = (\tau+\gamma)F_1'(\tau),
\label{relff1}
\ee
\be
\Psi(\tau) = (\tau+\gamma)\Psi_1(\tau)
\label{relpsipsi1}
\ee
with
\be
{\cal A} = (\lambda+\gamma)g_\lambda \frac{\partial}{\partial
\lambda} + (\mu+\gamma)g_\mu \frac{\partial}{\partial \mu} +
(\nu+\gamma)g_\nu \frac{\partial}{\partial \nu}
\ee
and ${\cal I}$ is the identity operator.

\section{The triaxial density as a function of $\Psi(\tau)$}

By noting that $H[\tau_1,\tau_1,\tau_2,\tau_3] =
\frac{d}{d\tau_1}H[\tau_1,\tau_2,\tau_3]$, we have
\be
H[\lambda,\mu,\nu,\lambda,\mu,\nu] =
\frac{\partial}{\partial \lambda}\frac{\partial}{\partial
\mu}\frac{\partial}{\partial \nu}H[\lambda,\mu,\nu].
\label{deffuncg}
\ee
Let $H(\tau)$ be the function 
\be
H(\tau) = (\tau+\alpha)^2(\tau+\beta)^2\Psi(\tau).
\label{deffuncgtau}
\ee
Then Eq.~(\ref{deffuncg}) can be written
\be
H[\lambda,\mu,\nu,\lambda,\mu,\nu] = 
\ee
\begin{displaymath}
\quad \frac{\partial}{\partial\lambda}\frac{\partial}{\partial\mu}\frac{\partial}{\partial\nu}
\left[ (\lambda+\alpha)(\lambda+\beta)\Psi(\lambda)g_\lambda \right] + (cyc) = 
\end{displaymath}
\be
\quad \frac{\partial}{\partial \lambda}\left[ (\lambda+\alpha) (\lambda+\beta) \Psi(\lambda)
\frac{\partial}{\partial \mu}\frac{\partial}{\partial \nu} g_\lambda \right] + (cyc),
\label{divdif1}
\ee
or, with Eq.~(\ref{gtaudef}),
\be
H[\lambda,\mu,\nu,\lambda,\mu,\nu] = g_\lambda^2 \Psi'(\lambda) +
2g_\lambda \Psi(\lambda) \frac{\partial g_\lambda}{\partial \lambda}  + (cyc).
\label{divdif2}
\ee
Using Eq.~(\ref{sumgtau}), we have
\be
\frac{\partial g_\lambda}{\partial \lambda} = \frac{g_\mu}{(\lambda-\mu)} +
\frac{g_\nu}{(\lambda-\nu)} \quad (cyc).
\label{derivg}
\ee
Substitution of Eq.~(\ref{derivg}) in Eq.~(\ref{divdif2}) and
comparison with Eq.~(\ref{densstackdef}) prove that the density $\rho$
can be expressed as the $5^{\rm th}$-order divided difference
\be
\rho(\lambda,\mu,\nu) = H[\lambda,\mu,\nu,\lambda,\mu,\nu]
\label{densdivdif}
\ee
with $H(\tau)$ given in Eq.~(\ref{deffuncgtau}).  By specifying a
function $\Psi(\tau)$ or the density profile on the $z$-axis
$\Psi'(\tau)$, one can easily calculate the density with
Eq.~(\ref{densdivdif}).  One can notice that any $4^{\rm th}$-order
polynomial added to the function $H(\tau)$ yields the same $5^{\rm
th}$-order divided difference as $H(\tau)$.

\section{The triaxial density as a function of $F(\tau)$}

One can also calculate the density from the potential by specifying
the function $F(\tau)$ instead of $\Psi(\tau)$. We show that the
triaxial density can be written as a $5^{\rm th}$-order divided
difference of a function $R(\tau)$ that depends only on $F(\tau)$.  

By direct application of Poisson's equation, using Eqs.~(\ref{defv1}),
(\ref{defnabla}) and (\ref{defdivdif3}), we can write the density
$\rho_1$ as
\be
\rho_1(\lambda,\mu,\nu) = R_1[\lambda,\mu,\nu] 
\label{dens1}
\ee
with
\bea
\pi G R_1(\tau) &=& a(\tau)F_1''(\tau) + \frac{1}{2}
a'(\tau)F_1'(\tau)\\ 
&=& \frac{a(\tau)}{\sqrt{|a(\tau)|}} \frac{d}{d
\tau} [\sqrt{|a(\tau)|}F_1'(\tau)].
\eea

Now we prove that the density $\rho$ can be written as the $5^{\rm
th}$-order divided difference
\be
\rho(\lambda,\mu,\nu) = R[\lambda,\mu,\nu,\lambda,\mu,\nu]
\label{densdivdifr2}
\ee
where 
\be
R(\tau) = a(\tau)R_1(\tau).
\label{deffuncrtau}
\ee
Using Eq.~(\ref{deffuncg}), we can write
\bea
\nonumber
R[\lambda,\mu,\nu,\lambda,\mu,\nu] &=&
\frac{\partial}{\partial\lambda} \frac{\partial}{\partial\mu}
\frac{\partial}{\partial\nu}\left[  
\frac{a(\lambda)R_1(\lambda)}{(\lambda-\mu)(\lambda-\nu)} + (cyc)\right]
\\
\nonumber
 & & \\
&=& \frac{\partial}{\partial\lambda} \left[ 
\frac{a(\lambda)R_1(\lambda)}{(\lambda-\mu)^2(\lambda-\nu)^2} \right] + (cyc).
\eea
Since
\be
\frac{a(\lambda)}{(\lambda-\mu)(\lambda-\nu)} =
(\lambda+\gamma)g_\lambda \quad (cyc),
\ee
we obtain
\be
R[\lambda,\mu,\nu,\lambda,\mu,\nu] = (\lambda+\gamma)g_\lambda
\frac{\partial}{\partial\lambda}
\left[ \frac{R_1(\lambda)}{(\lambda-\mu)(\lambda-\nu)}\right] +
\label{expressionr2}
\ee
\begin{displaymath}
\quad \quad \frac{R_1(\lambda)}{(\lambda-\mu)(\lambda-\nu)}
\frac{\partial}{\partial\lambda} [(\lambda+\gamma)g_\lambda ] + (cyc).
\end{displaymath}
With Eqs.~(\ref{sumgtau}) and (\ref{derivg}), one can establish the
identity
\be
\frac{\partial}{\partial \lambda} [(\lambda+\gamma)g_\lambda] =
1 + \frac{\mu+\gamma}{\lambda-\mu}g_\mu +
\frac{\nu+\gamma}{\lambda-\nu}g_\nu \quad (cyc),
\label{derivaux}
\ee
which, upon substitution of Eq.~(\ref{derivaux}) in
Eq.~(\ref{expressionr2}), yields
\be
R[\lambda,\mu,\nu,\lambda,\mu,\nu] = ({\cal A} + {\cal
I})R_1[\lambda,\mu,\nu] = ({\cal A} + {\cal I})\rho_1,
\ee
identical to Eq.~(\ref{relrhorho1}), q.e.d.

Thus, with Eq.~(\ref{relff1}), the density $\rho$ can be written as
the $5^{\rm th}$-order divided difference (Eq.~(\ref{densdivdifr2}))
with
\be
\pi G R(\tau) = |a(\tau)|^{\frac{3}{2}} \frac{d}{d
\tau} \left[ \frac{\sqrt{|a(\tau)|}}{\tau+\gamma} F(\tau) \right]
\label{funcrtau}
\ee
and $a(\tau)$ is given in Eq.~(\ref{defatau}).  The function $R(\tau)$
differs from the function $H(\tau)$ only by a polynomial function of
the $4^{\rm th}$-order in $\tau$.

% -----------------------------------------------------------------------

% -----------------------------------------------------------------------

\end{document}